\documentclass[aps,pra,showpacs,preprint]{revtex4}
\usepackage{graphicx}
\usepackage{amsmath,amssymb,amsfonts}
\usepackage{dcolumn}
\usepackage[english]{babel}
\usepackage{epsfig}

\begin{document}
\title{One- and two-dimensional Coulomb Green's function matrices in
 parabolic Sturmians basis}
\author{ S. A. Zaytsev}
\email[E-mail: ]{zaytsev@fizika.khstu.ru} \affiliation{Pacific
National University, Khabarovsk, 680035, Russia}

\begin{abstract}
One- and two-dimensional operators which originate from the
asymptotic form of the three-body Coulomb wave equation in parabolic
coordinates are treated within the context of square integrable
basis set. The matrix representations of Green's functions
corresponding to these operators are obtained.
\end{abstract}
\pacs{02.30Gp, 03.65Ca, 03.65.Nk} \maketitle

\section{Introduction}
The three Coulomb (3C) wave functions \cite{C31,C32} have been
introduced to approximate the three-body continuum Coulomb wave
function. The three-body Coulomb wave equation is asymptotically
separable in terms of parabolic coordinates introduced by Klar
\cite{Klar}
\begin{equation}\label{xieta}
    \xi_j = r_{ls}+\hat{\bf k}_{ls}\cdot {\bf r}_{ls}, \quad
    \eta_j = r_{ls}-\hat{\bf k}_{ls}\cdot {\bf r}_{ls},
\end{equation}
where ${\bf r}_{ls}$ and ${\bf k}_{ls}$, $l,\, s=1,\, 2,\, 3$ and $l
\ne s$, are the relative coordinate and momentum vectors between the
particles $l$ and $s$. The 3C wave functions are proportional to
solutions $\Psi_{3C}$ of the asymptotic equation \cite{Klar} (atomic
units are used):
\begin{equation}\label{AE}
    \left\{\sum_{j=1}^3 \frac{1}{\mu_{ls}(\xi_j+\eta_j)}
    \left[ \hat{h}_{\xi_j}+\hat{h}_{\eta_j}+2k_{ls}t_{ls}\right]
     \right\}\Psi_{3C}=0,
\end{equation}
where $t_{ls}=\frac{Z_l Z_s \mu_{ls}}{k_{ls}}$, $\mu_{ls}=\frac{m_l
m_s}{m_l+m_s}$ are the reduced masses. Here the operators
$\hat{h}_{\xi_j}$ and $\hat{h}_{\eta_j}$ are defined by
\begin{eqnarray}\label{hxi}
  \hat{h}_{\xi_j} &=& -2\left(\frac{\partial}{\partial\xi_j}\xi_j
    \frac{\partial}{\partial\xi_j}+ik_{ls}\xi_j
    \frac{\partial}{\partial\xi_j}\right),\\
  \label{heta}
  \hat{h}_{\eta_j} &=& -2\left(\frac{\partial}{\partial\eta_j}\eta_j
    \frac{\partial}{\partial\eta_j}-ik_{ls}\eta_j
    \frac{\partial}{\partial\eta_j}\right).
\end{eqnarray}
A solution $\Psi_{3C}$ with pure outgoing behavior can be written as
a product of three two-body Coulomb wave functions:
\begin{equation}\label{F3C}
    \Psi_{3C}=\prod \limits _{j=1}^3 u_j(\xi_j)
\end{equation}
with
\begin{equation}\label{outgoing}
    u_j(\xi_j)={_1 F _1(it_{ls},\, 1;\; -ik_{ls}\xi_j)}.
\end{equation}
The solutions (\ref{F3C}) are correct only when all particles are
far from each other. In the last years many works have been
published with different proposal to improve the simply 3C (see
\cite{C3mod1,C3mod2,MKP} and references therein). In this paper we
seek to consider the possibilities for computing the three-body
continuum Coulomb wave function which are afforded by expansion in a
set of square-integrable functions. In principle, the short-range
part of the Hamiltonian, which contains mixed second derivative
terms in the expression for the kinetic energy \cite{Klar}, can be
approximated by a finite order matrix. The long-range part
(\ref{AE}) of the Hamiltonian consists of two-dimensional operators
\begin{equation}\label{hhZj}
    \hat{\mathfrak{h}}_j=\hat{h}_{\xi_j}+\hat{h}_{\eta_j}+2k_{ls}t_{ls}
     +C_j(\xi_j+\eta_j),
\end{equation}
which are related by the constraint \cite{Klar}
\begin{equation}\label{Cc}
    C_1+C_2+C_3=0.
\end{equation}
To treat the one-dimensional operators
\begin{equation}\label{hxihetaC}
  \hat{h}_{\xi}+2kt +C\xi, \quad  \hat{h}_{\eta}+2kt +C\eta
\end{equation}
within the context of square integrable basis set, the $J$-matrix
method \cite{Jmtx1,Jmtx2} or the tools of the ``Tridiagonal
Physics'' program (see \cite{TDP} and reference therein) can be
employed. In the framework of these methods a special basis set is
used which supports an infinite tridiagonal matrix representation of
the operator. Thus the resulting three-term recursion is
analytically solved.

In particular, in this paper we obtain matrices of Green's functions
(resolvent) corresponding to the one-dimensional operators
(\ref{hxihetaC}) and the two-dimensional operator
\begin{equation}\label{hhZ}
    \hat{\mathfrak{h}}=\hat{h}_{\xi}+\hat{h}_{\eta}+2kt_0 +C(\xi+\eta)
\end{equation}
in the set of parabolic Sturmian functions,
\begin{equation}\label{BBF}
  \phi_{n_1\,n_2}(\xi,\,\eta)=\varphi_{n1}(\xi)\,\varphi_{n2}(\eta),
\end{equation}
\begin{equation}\label{BF}
  \varphi_n(x)=\sqrt{2b} e^{-bx}L_{n}(2bx),
\end{equation}
where $b$ is the scale parameter. This basis set have been used in
the analysis of the Coulomb potential within the parabolic
formulation of the $J$-matrix method  \cite{Ojha1}. The basis
functions (\ref{BF}) are orthonormal:
\begin{equation}\label{OBF}
    \int\limits _{0}^{\infty}dx\, \varphi_{n}(x)\,\varphi_{m}(x) =
    \delta_{n\,m}.
\end{equation}
In Sec.~II using the tridiagonal matrix representations of
one-dimensional operators (\ref{hxihetaC}) in the bases (\ref{BF}),
we construct the corresponding Green's function matrices. In this
case we do not seek to determine the Green's matrices uniquely. In
Sec.~III the weight function is obtained for the orthogonal
polynomials satisfying the three-term recurrence relation. The
two-dimensional Green's function matrix elements are expressed as
convolution of one-dimensional Green's matrix elements in Sec.~III.
An orthogonality relation employed in the two-dimensional Green's
matrix construction is derived in the Appendix.

\section{One-dimensional Coulomb Green's function matrices}
\section*{a) $C=0$}
The matrix representation ${\bf h}_{\xi}+2kt{\bf I}_{\xi}$ [${\bf
I}_{\xi}$ is the unit matrix] of the operator $\hat{h}_{\xi}+2 k t$
in the basis set $\left\{\varphi_n(\xi)\right\}_{n=0}^{\infty}$
(\ref{BF}) is tridiagonal
\begin{equation} \label{hnum}
 {\bf h}_{\xi}+2kt{\bf I}_{\xi} = \left(
 \begin{array}{ccccccc}
  b_0 & d_1 & & &   \\
  a_1 & b_1 & d_2 & & \mbox{\Large $0$} &  \\
      & a_2 & b_2 & d_3 & &  \\
  && a_3& \times & \times &  \\
  &\mbox{\Large $0$} &&\times & \times & \times\\
 \end{array}
\right),
\end{equation}
where
\begin{equation}\label{ab1}
 \begin{array}{c}
  b_n=(b+ik)+2bn+2kt,\\[2mm]
  a_n = (b-ik)n, \quad d_n = (b+ik)n.\\
 \end{array}
\end{equation}
To construct the Green's function matrix ${\bf g}_{\xi}$, which is
matrix inverse of the infinite tridiagonal matrix (\ref{hnum}),
consider the three-term recurrence relation
\begin{equation}\label{TRR1}
    a_n\,w_{n-1}+b_n\,w_n+d_{n+1}\,w_{n+1}=0,
    \quad n \ge 1.
\end{equation}
It can be easily verified that
\begin{equation}\label{pn}
    p_n(t; \; \zeta) = \frac{(-1)^n}{n!}\frac{\Gamma(n+1-it)}{\Gamma(1-it)}\;
    {_2F_1(-n,\,it;\;-n+it;\;\zeta)},
\end{equation}
where $\zeta=\frac{b-ik}{b+ik}$, is the ``regular'' solution of
(\ref{TRR1}) which satisfies the initial conditions:
\begin{equation}\label{Bp}
    p_0(t;\; \zeta)=1, \quad
    b_0\,p_0(t;\; \zeta)+d_1\,p_1(t;\; \zeta)=0.
\end{equation}
Suffice it to say that apart from the factor
$\sqrt{\frac{2}{b}}\left(\frac{\zeta+1}{2} \right)^{i t}$, $p_n$ is
the coefficient of the $n$th basis function $\varphi_n(\xi)$
(\ref{BF}) in the expansion of $u(\xi)$ (\ref{outgoing}). Note that
$p_n$ are polynomials of degree $n$ in $t$.

The second solution $q_n$ of the recursion (\ref{TRR1}) can be
obtained from the condition that $q_n$ satisfies the same
differential equation as the $p_n$ \cite{Jmtx2}. In other words, if
$p_n \sim {_2F_1(a,\,b;\;c;\;z)}$, then $q_n \sim
z^{1-c}{_2F_1(a-c+1,\,b-c+1;\;2-c;\;z)}$. It is readily verified
that an appropriate $q_n(t;\; \zeta)$ is
\begin{equation}\label{qn}
 \begin{array}{c}
    q_n(t;\; \zeta) = -\frac{n!\Gamma(1-it)}{\Gamma(n+2-it)}\,(-\zeta)^{n+1}\,
    {_2F_1(1-it,\,n+1;\;n+2-it;\;\zeta)}\\[3mm]
    \hphantom{q_n(t) = }=-\frac{n!\Gamma(1-it)}{\Gamma(n+2-it)}\,
     \left(\frac{\zeta}{\zeta-1}\right)^{n+1}\,
    {_2F_1(n+1,\,n+1;\;n+2-it;\;\frac{\zeta}{\zeta-1})}\\
 \end{array}
\end{equation}
This satisfies the initial condition
\begin{equation}\label{Bq}
    b_0\,q_0(t;\; \zeta)+d_1\,q_1(t;\; \zeta)=b-ik.
\end{equation}

Multiplying $\left[{\bf h}_{\xi}+2kt{\bf I}_{\xi}\right]$ by the
diagonal matrix ${\bf Z}=\left[1,\, \zeta^{-1}, \, \ldots, \,
\zeta^{-n}, \, \ldots \right]$, we obtain the the symmetric
tridiagonal matrix ${\bf T}$:
\begin{equation}\label{hnumt}
{\bf T} = {\bf Z}\,\left[{\bf h}_{\xi}+2kt{\bf I}_{\xi}\right]=
\left(
 \begin{array}{ccccccc}
  \beta_0 & \alpha_1 & & &   \\
  \alpha_1 & \beta_1 & \alpha_2 & & \mbox{\Large $0$} &  \\
      & \alpha_2 & \beta_2 & \alpha_3 & &  \\
  && \alpha_3& \times & \times &  \\
  &\mbox{\Large $0$} &&\times & \times & \times\\
 \end{array}
\right)
\end{equation}
with nonzero elements
\begin{equation}\label{ab0}
\beta_n=b_n/\zeta^{n}, \quad \alpha_n = d_n/\zeta^{n-1}.
\end{equation}
Notice that $p_n$ and $q_n$ satisfy the three term recurrence
relation
\begin{equation}\label{TRR2}
    \alpha_n\,w_{n-1}+\beta_{n}\,w_n+\alpha_{n+1}\,w_{n+1}=0, \quad
    n \ge 1.
\end{equation}
Thus, to invert the symmetric tridiagonal matrix ${\bf T}$, one can
draw on the standard method \cite{Case,Simon}. Namely, the elements
of a Green's matrix ${\bf g}_T$, which is the matrix inverse to
${\bf T}$, can be determined by
\begin{equation}\label{gnmt}
    g^{T}_{nm}(t)=\frac{p_{\nu}(t;\; \zeta)\,q_{\mu}(t;\; \zeta)}
    {W(q,\,p)},\quad
    \nu=\min(n,m), \quad \mu=\max(n,m),
\end{equation}
where the Wronskian $W$ is defined as
\begin{equation}\label{Wrn}
 \begin{array}{c}
    W(q, \, p) = \alpha_n\left[q_{n}(t;\; \zeta)\,p_{n-1}(t;\; \zeta)
    -q_{n-1}(t;\; \zeta)\,p_{n}(t;\; \zeta)
    \right]\\[2mm]
    \qquad \qquad \qquad =b-ik=-2ik\left(\frac{\zeta}{\zeta-1}\right).
 \end{array}
\end{equation}

The Green's matrix ${\bf g}_{\xi}$=$\left[{\bf h}_{\xi}+2kt{\bf
I}_{\xi}\right]^{-1}$ is related to ${\bf g}_T$: ${\bf g}_{\xi}={\bf
g}_T\,{\bf Z}$. Therefore, we can express the matrix ${\bf g}_{\xi}$
elements in the form
\begin{equation}\label{gnm}
    g^{\xi}_{nm}(t)=\frac{i}{2k}
    \left(\frac{\zeta-1}{\zeta}\right)\frac{1}{\zeta^m}\,
    p_{\nu}(t;\; \zeta)\,q_{\mu}(t;\; \zeta).
\end{equation}

From (\ref{hxi}) and (\ref{heta}) it follows that the Green's matrix
${\bf g}_{\eta}$=$\left[{\bf h}_{\eta}+2kt{\bf
I}_{\eta}\right]^{-1}$ and ${\bf g}_{\xi}$ are complex conjugates
(for real $k$ and $t$):
\begin{equation}\label{gxgt}
    g^{\eta}_{n\,m}(t)=\left(g^{\xi}_{n\,m}(t) \right)^*.
\end{equation}

Note that there is an ambiguity in determining the matrix ${\bf
g}_{\xi}$, since the solution $q_n(t;\; \zeta)$ is not unique:
\begin{equation}\label{qt}
    \widetilde{q}_n(t;\; \zeta)=q_n(t;\; \zeta)+y(t)p_n(t;\; \zeta),
\end{equation}
where $y(t)$ is an arbitrary function of $t$, also satisfies
(\ref{TRR1}).

\section*{b) $C \neq 0$}
It is not difficult to convince oneself that the differential
equation
\begin{equation}
\left[\hat{h}_{\xi}+2kt+C\xi\right]u(\xi)=0
\end{equation}
is satisfied by the function
\begin{equation}\label{u2}
    u(\xi)=e^{\lambda \xi}\, {_1 F_1} \left(i \tau,\, 1;\;
    -i\gamma\xi\right),
\end{equation}
where
\begin{equation}\label{lg}
    \lambda=ik\frac{\sqrt{1-\frac{2C}{k^2}}-1}{2}, \;
     \tau = \frac{2t+i\left(1-\sqrt{1-\frac{2C}{k^2}}\right)}{2\sqrt{1-\frac{2C}{k^2}}},\;
      \gamma=k\sqrt{1-\frac{2C}{k^2}}.
\end{equation}
Clearly, in the limit $C=0$ the function (\ref{u2}) reduces to
(\ref{outgoing}).

Because the operator $\xi$, evaluated in the basis (\ref{BF}), has
the symmetric tridiagonal form
\begin{equation}\label{Q}
    Q_{n,\, n'}=\left\{
    \begin{array}{lr}
     -\frac{1}{2b}\,n, & n'=n-1,\\
     \frac{1}{2b}\,(2n+1), & n'=n,\\
     -\frac{1}{2b}\,(n+1), & n'=n+1,\\
    \end{array}
    \right.
\end{equation}
the matrix representation of the operator $\hat{h}_{\xi}+2kt+C\xi$
is also tridiagonal. Thus,
\begin{equation}\label{hx}
    {\bf h}_{\xi}+2kt{\bf I}_{\xi}+C{\bf Q}_{\xi}=
    \left(
 \begin{array}{ccccccc}
  b_0 & d_1 & & &   \\
  a_1 & b_1 & d_2 & & \mbox{\Large $0$} &  \\
      & a_2 & b_2 & d_3 & &  \\
  && a_3& \times & \times &  \\
  &\mbox{\Large $0$} &&\times & \times & \times\\
 \end{array}
\right),
\end{equation}
where
\begin{equation}\label{ab2}
 \begin{array}{c}
  b_n=(b+\frac{C}{2b}+ik)+2(b+\frac{C}{2b})n+2kt,\\[2mm]
  a_n = (b-\frac{C}{2b}-ik)n,\quad d_n = (b-\frac{C}{2b}+ik)n.\\
 \end{array}
\end{equation}

Then, if we consider the coefficients of the regular solution
(\ref{u2}) expansion in the basis set (\ref{BF}), we obtain that
\begin{equation}\label{sn}
    s_n(t;\;C) = \theta^n\,p_n(\tau;\; \zeta),
\end{equation}
where
\begin{equation}\label{ttz}
     \theta=\frac{b+\lambda}{b-\lambda},\quad \zeta=\frac{b+\frac{C}{2b}
     -ik\sqrt{1-\frac{2C}{k^2}}}
    {b+\frac{C}{2b}+ik\sqrt{1-\frac{2C}{k^2}}},
\end{equation}
is the solution of the three term recurrence relation
\begin{equation}\label{TRR3}
    a_n\,w_{n-1}+b_n\,w_n+d_{n+1}\,w_{n+1}=0,
    \quad n \ge 1.
\end{equation}
It is easy to verify that a second solution of Eq. (\ref{TRR3}) can
be expressed in the form
\begin{equation}\label{cn}
    c_n(t;\;C) =\theta^{n+1}\,q_n(\tau;\;
    \zeta).
\end{equation}

Notice that the matrix ${\bf h}_{\xi}+2kt{\bf I}_{\xi}+C{\bf
Q}_{\xi}$ inversion procedure is simplified if we introduce the
symmetric tridiagonal matrix ${\bf T}$:
\begin{equation}\label{hnumt2}
{\bf T} = {\bf Z}\,\left[{\bf h}_{\xi}+2kt{\bf I}_{\xi}+C{\bf
Q}_{\xi}\right]= \left(
 \begin{array}{ccccccc}
  \beta_0 & \alpha_1 & & &   \\
  \alpha_1 & \beta_1 & \alpha_2 & & \mbox{\Large $0$} &  \\
      & \alpha_2 & \beta_2 & \alpha_3 & &  \\
  && \alpha_3& \times & \times &  \\
  &\mbox{\Large $0$} &&\times & \times & \times\\
 \end{array}
\right).
\end{equation}
Here ${\bf Z}$ is the diagonal matrix: ${\bf Z}=\left[1,\,
\chi^{-1}, \, \ldots, \, \chi^{-n}, \,  \ldots \right]$,
$\chi=\frac{b-\frac{C}{2b}-ik}{b-\frac{C}{2b}+ik}$. The elements
$\beta_n$ and $\alpha_n$ are given by
\begin{equation}\label{ab3}
    \beta_n=b_n/\chi^n, \quad \alpha_n=d_n/\chi^{n-1}.
\end{equation}
Then, it is easy to check that $s_n$ and $c_n$ also satisfy the
recursion equation
\begin{equation}\label{TRR4}
    \alpha_n\,w_{n-1}+\beta_{n}\,w_n+\alpha_{n+1}\,w_{n+1}=0, \quad
    n \ge 1,
\end{equation}
and the Wronskian
\begin{equation}\label{W2}
    W(c,s)=\alpha_n\left[c_{n}(t;\; C)s_{n-1}(t;\; C)-c_{n-1}(t;\; C)s_{n}(t;\; C)
    \right]
\end{equation}
is independent of $n$. Namely,
\begin{equation}\label{W3}
W(c,s)=b-\frac{C}{2b}-ik=-2ik\left(\frac{\chi}{\chi-1}\right).
\end{equation}
Finally, given the two linearly independent solutions $s_n$ and
$c_n$ of Eq. (\ref{TRR3}), we can express the Green's matrix ${\bf
g}_{\xi}=[{\bf h}_{\xi}+2kt{\bf I}_{\xi}+C{\bf Q}_{\xi}]^{-1}$
elements in the form
\begin{equation}\label{gnm2}
    g^{\xi}_{nm}(t;\; C)=\frac{s_{\nu}(t;\;C)\,c_{\mu}(t;\;C)}{W(c,s)}\,\chi^{-m}
    =\frac{i}{2k}
    \left(\frac{\chi-1}{\chi}\right)\frac{\theta^{n+m+1}}{\chi^m}\,
    p_{\nu}(\tau;\; \zeta)\,q_{\mu}(\tau;\; \zeta).
\end{equation}

Clearly, the Green's matrix ${\bf g}_{\eta}$=$\left[{\bf
h}_{\eta}+2kt{\bf I}_{\eta}+C{\bf Q}_{\eta}\right]^{-1}$ is obtained
from ${\bf g}_{\xi}$ by replacing $t \rightarrow -t$, $k \rightarrow
-k$ [this leaves $kt$ unchanged].

\section{Orthogonal polynomials $p_n$}
In this section we obtain the weight function with respect to which
the polynomials $p_n$ (\ref{pn}) are orthonormal.

The appropriate Kummer's relation [(15.3.7) in Ref.
\cite{Abramowitz}] expresses the solution $q_n$ as
\begin{equation}\label{qnp}
  \begin{array}{l}
    q_n(t;\; \zeta) =-\frac{n!\Gamma(-n-it)}{\Gamma(1-it)}\;
    {_2F_1(it,\,n+1;\;n+1+it;\;{\zeta}^{-1})}-\\[2mm]
    \hfill -\Gamma(1-it)\Gamma(it)(-\zeta)^{it}\,p_n(t;\; \zeta),\\
  \end{array}
\end{equation}
\begin{equation}\label{acon}
    \left|\arg(-\zeta)\right| < \pi.
\end{equation}
Then introduce the function
\begin{equation}\label{rho1}
    \rho(t;\; \zeta) = \frac{\Gamma(1-it)\Gamma(it)(-\zeta)^{it}}{2 \pi i}
\end{equation}
and evaluate the integral
\begin{equation}\label{I1}
    \mathcal{I}(\zeta) = \int \limits_{\mathcal{C}}dt\, \rho(t;\;\zeta),
\end{equation}
where $\mathcal{C}$ runs along real axis, except for an
infinitesimal indentation on the underside of the pole $t=0$; see
Fig.~1.

To evaluate $\mathcal{I}(\zeta)$ for $\left|\zeta\right|>1$, change
the variable of integration to $s$: $t=\frac{-is}{s+i}$,
transforming the path of integration into the circle $\mathcal{C}_1$
in the complex $s$-plane; see Fig.~2. Thus the interior of
$\mathcal{C}_1$ corresponds to the upper half of the $t$-plane and
hence the integral
\begin{equation}\label{I2}
    \mathcal{I}(\zeta) = \int \limits_{\mathcal{C}_1}\frac{ds}{(s+i)^2}\,
    \rho_1(s;\;\zeta),
\end{equation}
where
$\rho(t;\;\zeta)=\rho(\frac{-is}{s+i};\;\zeta)=\rho_1(s;\;\zeta)$,
is reduced to the sum of the residues at the poles [corresponding to
$t_n=in, \, n=0,1,\ldots$] inside the circle contour
$\mathcal{C}_1$:
\begin{equation}\label{I3}
    \mathcal{I}(\zeta) = i \sum
    \limits_{n=0}^{\infty}\frac{1}{\zeta^n},
\end{equation}
i. e.
\begin{equation}\label{I4}
    \mathcal{I}(\zeta) = \frac{i \zeta}{1-\zeta}.
\end{equation}

In the case $\left|\zeta \right|<1 $ it is convenient to change
variables from $t$ to $s$: $t=\frac{i s}{s+i}$. The integration
contour $\mathcal{C}$ is transformed into the circle $\mathcal{C}_2$
[see Fig.~3] and the lower half of the $t$-plane is mapped onto the
interior of $\mathcal{C}_2$. Then introducing $\rho_2(s;\;\zeta) =
\rho(\frac{i s}{s+i};\;\zeta)=\rho(t;\;\zeta)$, we obtain
\begin{equation}\label{I5}
    \mathcal{I}(\zeta) = \int \limits_{\mathcal{C}_2}\frac{-ds}{(s+i)^2}\,
    \rho_2(s;\;\zeta)=i \sum
    \limits_{n=1}^{\infty}\zeta^n=\frac{i \zeta}{1-\zeta}.
\end{equation}

Note that for $Im(\zeta)\neq 0$ (\ref{acon}) implies that
\begin{equation}\label{rho2}
    \rho(t;\; \zeta)=\left\{
    \begin{array}{lr}
     \frac{\displaystyle \zeta^{i t}}
     {\displaystyle 1-e^{2 \pi t}}, & -\pi < \arg(\zeta)
     <0,\\[4mm]
     \frac{\displaystyle \zeta^{i t}}
     {\displaystyle e^{-2 \pi t}-1}, & 0 <
     \arg(\zeta) < \pi.\\
    \end{array}
    \right.
\end{equation}
Using the Sokhotsky formula we obtain
\begin{equation}\label{I6}
\mathcal{I}(\zeta) =
\mathcal{P}\!\!\!\int\limits_{-\infty}^{\infty}dt\,
    \rho(t;\; \zeta)-\frac{i}{2}.
\end{equation}
In this case (\ref{rho2}) provides the convergence of the
principal-value integral on the right hand side of (\ref{I6}).

Clearly, integrals
\begin{equation}\label{I7}
    \int \limits _{\mathcal{C}} dt\, \rho(t;\;\zeta)\,p_n(t;\;\zeta)\,p_m(t;\;\zeta)
\end{equation}
can be expanded in terms of derivatives of (\ref{I1}). It is now
easy to obtain that
\begin{equation}\label{Orth1}
    \frac{i}{\zeta^n}\left(\frac{\zeta-1}{\zeta}\right)
    \int \limits _{\mathcal{C}} dt \rho(t;\;\zeta)\,
    p_n(t;\;\zeta)\,p_m(t;\;\zeta)=\delta_{n\,m},
\end{equation}
i. e. $\rho$  (\ref{rho1}) is the weight function for the orthogonal
polynomials $p_n$ (\ref{pn}).

For $Im(\zeta)\neq 0$ we can rewrite the orthonormality relation
(\ref{Orth1}) as
\begin{equation}\label{Orth2}
    \frac{i}{\zeta^n}\left(\frac{\zeta-1}{\zeta}\right)
    \left(\mathcal{P}\!\!\!\int\limits_{-\infty}^{\infty}dt\,
    \rho(t;\; \zeta)\,p_n(t;\; \zeta)\,p_{m}(t;\; \zeta)
     -\frac{i}{2}(-1)^{n+m}\right)=\delta_{n\,m},
\end{equation}
where the weight function $\rho$ in the form (\ref{rho2}) is used.
Note that in the case $C=0$ we have $\zeta=e^{i\varphi}$,
$\varphi<0$, and the weight function reduces to
\begin{equation}\label{rho0}
    \rho_0(t; \; \zeta) = \frac{e^{-\varphi t}}{1-e^{2 \pi t}}.
\end{equation}

It may be remarked that the polynomials $p_n$ are discrete analogues
of the charge parabolic Coulomb Sturmians introduced in
\cite{Gasaneo3}.

\section{Two-dimensional Green's function matrices}
In this section we obtain a matrix representation ${\bf G}(t_0;\;
C)$ of the Green's function (resolvent) of the operator (\ref{hhZ}).
Formally ${\bf G}(t_0;\; C)$ is the matrix inverse of the infinite
matrix
\begin{equation}\label{twoh}
    {\bf h}={\bf h}_{\xi}\otimes{\bf I}_{\eta}+{\bf I}_{\xi}\otimes{\bf h}_{\eta}
    +2kt_0{\bf I}_{\xi}\otimes{\bf I}_{\eta}
    +C\left({\bf Q}_{\xi}\otimes{\bf I}_{\eta}+{\bf I}_{\xi}\otimes{\bf
    Q}_{\eta}\right),
\end{equation}
i. e.
\begin{equation}\label{UnitI}
{\bf h}\,{\bf G}(t_0;\; C)={\bf I}_{\xi}\otimes{\bf I}_{\eta}.
\end{equation}

\section*{a) $C=0$}
The elements of ${\bf G}(t_0) \equiv {\bf G}(t_0;\; 0)$ can be
expressed as a convolution integral (see, e. g. \cite{Shakeshaft})
\begin{equation}\label{Conv}
G_{n_1\,n_2,\, m_1\,m_2}(t_0)= \int \limits _{\mathcal{C}_0} dt\,
    \tilde{g}^{\xi}_{n_1\,m_1}(t)\,
    \tilde{g}^{\eta}_{n_2\,m_2}(t_0-t),
\end{equation}
where the integrand contains functions $\tilde{g}^{\xi}_{n\,m}$ and
$\tilde{g}^{\eta}_{n\,m}$ that are proportional to the
one-dimensional Green's function matrix elements [the non-uniqueness
(\ref{qt}) of the solution $q_n$ is taken into account]:
\begin{equation}\label{gxi2}
    \tilde{g}^{\xi}_{n\,m}(t) = \frac{i}{2k}
    \left(\frac{\zeta-1}{\zeta}\right)\frac{1}{\zeta^m}\,
    p_{\nu}(t;\; \zeta)\left[A_{\xi}\,q_{\mu}(t;\; \zeta)
    +x_{\mu}(t)\,p_{\mu}(t;\; \zeta)\right]B_m
\end{equation}
and
\begin{equation}\label{geta2}
    \tilde{g}^{\eta}_{n\,m}(t) = \frac{i}{2k}(\zeta-1)\zeta^m\,
    p_{\nu}(-t;\; \zeta^{-1})\left[A_{\eta}\,q_{\mu}(-t;\; \zeta^{-1})
    +y_{\mu}(t)\,p_{\mu}(-t;\; \zeta^{-1})\right]D_m.
\end{equation}
Inserting (\ref{Conv})-(\ref{geta2}) into (\ref{UnitI}) then gives
the restriction on the factors $A_{\xi}$ and $A_{\eta}$, functions
$\left\{x_n(t)\right\}_{n=0}^{\infty}$ and
$\left\{y_n(t)\right\}_{n=0}^{\infty}$, diagonal matrices ${\bf
B}=[B_0, B_1, \, \ldots]$ and ${\bf D}=[D_0, D_1, \, \ldots]$, and
the path of integration $\mathcal{C}_0$, namely,
\begin{eqnarray}\label{Dlt1}
  A_{\eta}D_{m_2}\int \limits _{\mathcal{C}_0} dt\,
    \tilde{g}^{\xi}_{n_1\,m_1}(t) &=& U_{n_1\,m_2}\delta_{n_1\,m_1},\\
    \label{Dlt2}
  A_{\xi}B_{m_1}\int \limits
_{\mathcal{C}_0} dt\,
    \tilde{g}^{\eta}_{n_2\,m_2}(t_0-t) &=& V_{m_1\,n_2}\delta_{n_2\,m_2},\\
    \label{Dlt3}
  U_{n_1\,m_2}+V_{m_1\,n_2} &=& 1.
\end{eqnarray}

If we choose the contour $\mathcal{C}$ (see Fig.~1) as the path of
integration in (\ref{Conv}) [and (\ref{Dlt1}), (\ref{Dlt2})], we can
draw on the orthonormality relation (\ref{Orth1}) to determine the
rest of parameters. In this case we readily check that, in
particular, the set: $A_{\xi}=0$, $x_n(t)=2\pi i \rho_0(t;\;
\zeta)$, $A_{\eta}=\frac{k}{i \pi}$, $y_n(t)\equiv0$, $B_n=D_n=1$,
satisfies the conditions (\ref{Dlt1})-(\ref{Dlt3}), and hence the
elements $G_{n_1\,n_2,\, m_1\,m_2}$ can be expressed in the form
\begin{equation}\label{Conv2}
 \begin{array}{c}
G_{n_1\,n_2,\, m_1\,m_2}(t_0) =
\frac{i}{\zeta^{m_1}}\left(\frac{\zeta-1}{\zeta}\right)\left(
\mathcal{P}\!\!\!\int \limits _{-\infty}^{\infty} dt\,
    \rho_0(t;\; \zeta)p_{n_1}(t;\; \zeta)\,p_{m_1}(t;\; \zeta)\,
     g^{\eta}_{n_2\,m_2}(t_0-t)\right.\\[3mm]
     \hfill
     \left. - \frac{i}{2}(-1)^{n_1+m_1}\,g^{\eta}_{n_2\,m_2}(t_0)
     \vphantom{\int \limits _{-\infty}^{\infty}} \right).
 \end{array}
\end{equation}
Further, combining (\ref{Orth1}) with (\ref{Ignm}) and (\ref{gxgt}),
we can obtain another allowable sets of the parameters. For
instance, $A_{\xi}=1$, $x_n(t)=2\pi i \rho_0(t;\; \zeta)$,
$A_{\eta}=\frac{k}{i \pi}$, $y_n(t)\equiv 0$, $B_n=D_n=1$.

\section*{b) $C \neq 0$}
The elements of the matrix ${\bf G}(t_0;\; 0)$ may be written as the
convolution integral
\begin{equation}\label{Conv3}
G_{n_1\,n_2,\, m_1\,m_2}(t_0;\; C)= \int \limits _{\mathcal{C}_3}
d\tau \,
    \tilde{g}^{\xi}_{n_1\,m_1}(t;\; C)\,
    g^{\eta}_{n_2\,m_2}(t_0-t;\; C).
\end{equation}
Here, the non-uniqueness of only ${\bf g}_{\xi}$ is taken into
consideration for simplicity, i. e. $\tilde{g}^{\xi}_{n\,m}$ and
$g^{\eta}_{n\,m}$ are taken to be
\begin{equation}\label{gxi3}
    \tilde{g}^{\xi}_{n\,m}(t;\; C) = \frac{i}{2k}
    \left(\frac{\chi-1}{\chi}\right)\frac{\theta^{n+m+1}}{\chi^m}\,
    p_{\nu}(\tau;\; \zeta)\left[A_{\xi}\,q_{\mu}(\tau;\; \zeta)
    +x_{\mu}(\tau)\,p_{\mu}(\tau;\; \zeta)\right]B_m,
\end{equation}
\begin{equation}\label{geta3}
    g^{\eta}_{n\,m}(t_0-t;\; C) = \frac{i}{2k}
    \frac{(\chi-1)\chi^m}{\theta^{n+m+1}}\,
    p_{\nu}\left(\tau-\frac{t_0}{\sqrt{1-\frac{2C}{k^2}}};\; \zeta^{-1}\right)
    \,q_{\mu}\left(\tau-\frac{t_0}{\sqrt{1-\frac{2C}{k^2}}};\; \zeta^{-1}\right).
\end{equation}
The path of integration $\mathcal{C}_3$ in the complex $\tau$-plane
(see Fig.~4) is identical to the contour $\mathcal{C}$ in the
complex $t$-plane. The integration over $\tau$ (\ref{ttz}) is
performed on the assumption that $\tau$ and $C$ are independent of
one another.

Allowable parameters $A_{\xi}$, $x_n(\tau)$ and $B_m$ can be
determined by substitution (\ref{Conv3})-(\ref{geta3}) into
(\ref{UnitI}). For instance, putting $A_{\xi}=0$ and
$x_n(\tau)=2k\rho(\tau;\; \zeta)$, we obtain
$B_m=\frac{1}{\theta^{2m+1}}\frac{(\zeta-1)}{(\chi-1)}\left(\frac{\chi}
{\zeta}\right)^{m+1}$. Thus the matrix elements of the
two-dimensional Green's function can be written in the form
\begin{equation}\label{Conv4}
 \begin{array}{l}
G_{n_1\,n_2,\, m_1\,m_2}(t_0;\; C) =\\[3mm]
\hfill
=\frac{i}{\zeta^{m_1}}\left(\frac{\zeta-1}{\zeta}\right)\,\theta^{n_1-m_1}
\int \limits _{\mathcal{C}_3} d\tau\,
    \rho(\tau;\; \zeta)\,p_{n_1}(\tau;\; \zeta)\,p_{m_1}(\tau;\; \zeta)\,
     g_{n_2\,m_2}^{\eta}(t_0-t;\; C).
 \end{array}
\end{equation}

\appendix{}
\section{One useful orthogonality relation}
In this appendix we derive the orthogonality relation
\begin{equation}\label{Ignm}
    \frac{2 i k}{\pi}\int \limits _{-\infty}^{\infty}
    dt\, g^{\xi}_{n\,m}(t)
    =\delta_{n\,m}.
\end{equation}
Using the integral representation [Eq. (15.3.1) in Ref.
\cite{Abramowitz}] of the hypergeometric function in (\ref{qn}), we
can rewrite the integral on the left hand side of (\ref{Ignm}) in
the form
\begin{equation}\label{Ignm2}
    \frac{1}{\pi}\left(\frac{\zeta}{\zeta-1}\right)^{\mu}
    \frac{1}{\zeta^m}\,
    \int \limits_{0}^{1}dx \frac{x^{\mu}}
    {\left(1-x\frac{\zeta}{\zeta-1}\right)^{\mu+1}}
    \int \limits _{-\infty}^{\infty} dt\, p_{\nu}(t; \; \zeta)\,e^{-it\ln(1-x)}.
\end{equation}
Notice that the integral over $t$ in (\ref{Ignm2}) consists of
integrals
\begin{equation}\label{Ignm3}
    \int \limits _{-\infty}^{\infty} dt\, (-i t)^{\ell}\,e^{-it\ln(1-x)}
    =2\pi\left[(x-1)\frac{d}{dx}\right]^{\ell}\delta(x), \qquad \ell
    \le \nu \le \mu,
\end{equation}
i. e. is expressed in terms of derivatives of the delta function.
Further, introducing the function
\begin{equation}\label{fu}
    y_j(x)=\frac{x^{\mu}(x-1)^j}{\left(1-x\frac{\zeta}{\zeta-1}\right)^{\mu+1}},
\end{equation}
we obtain
\begin{equation}\label{Ignm4}
   \mathcal{I}_{j\,\ell}\equiv \int \limits_{0}^{1}dx \,
   y_j(x)\,\delta^{(\ell)}(x)=\left\{
   \begin{array}{lr}
    \frac12\,(-1)^{\ell}\,y_j^{(\ell)}(0)=\frac{\mu!}{2}\, (-1)^{\ell+j},& \ell=\mu,\\
    0, & \ell< \mu.\\
   \end{array}
   \right.
\end{equation}
From (\ref{Ignm4}) it follows that the integral in (\ref{Ignm2}) is
nonzero if $n=m$ $(=\nu=\mu)$. In this case the only contribution to
the integral comes from the leading term of the polynomial $p_n(t;\;
\zeta)$ [which is equal to
$\frac{1}{n!}\left(\zeta-1\right)^n(-it)^{n}$]. Thus, we obtain for
(\ref{Ignm2}):
\begin{equation}\label{Ignm5}
\frac{1}{\pi}
\left(\frac{\zeta}{\zeta-1}\right)^{n}\frac{1}{\zeta^n}\,
\frac{1}{n!}\left(\zeta-1\right)^n\,2\pi\, \mathcal{I}_{n\,n}=1.
\end{equation}

\begin{figure*}[ht]
\centerline{\psfig{figure=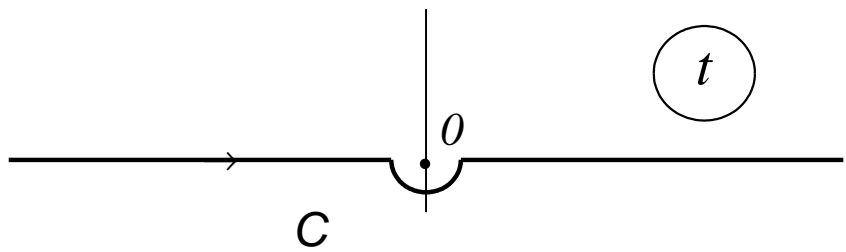,width=1\textwidth}} \caption{The
path of integration $\mathcal{C}$ in (\ref{I1}).}
\end{figure*}

\newpage
\begin{figure*}[ht]
\centerline{\psfig{figure=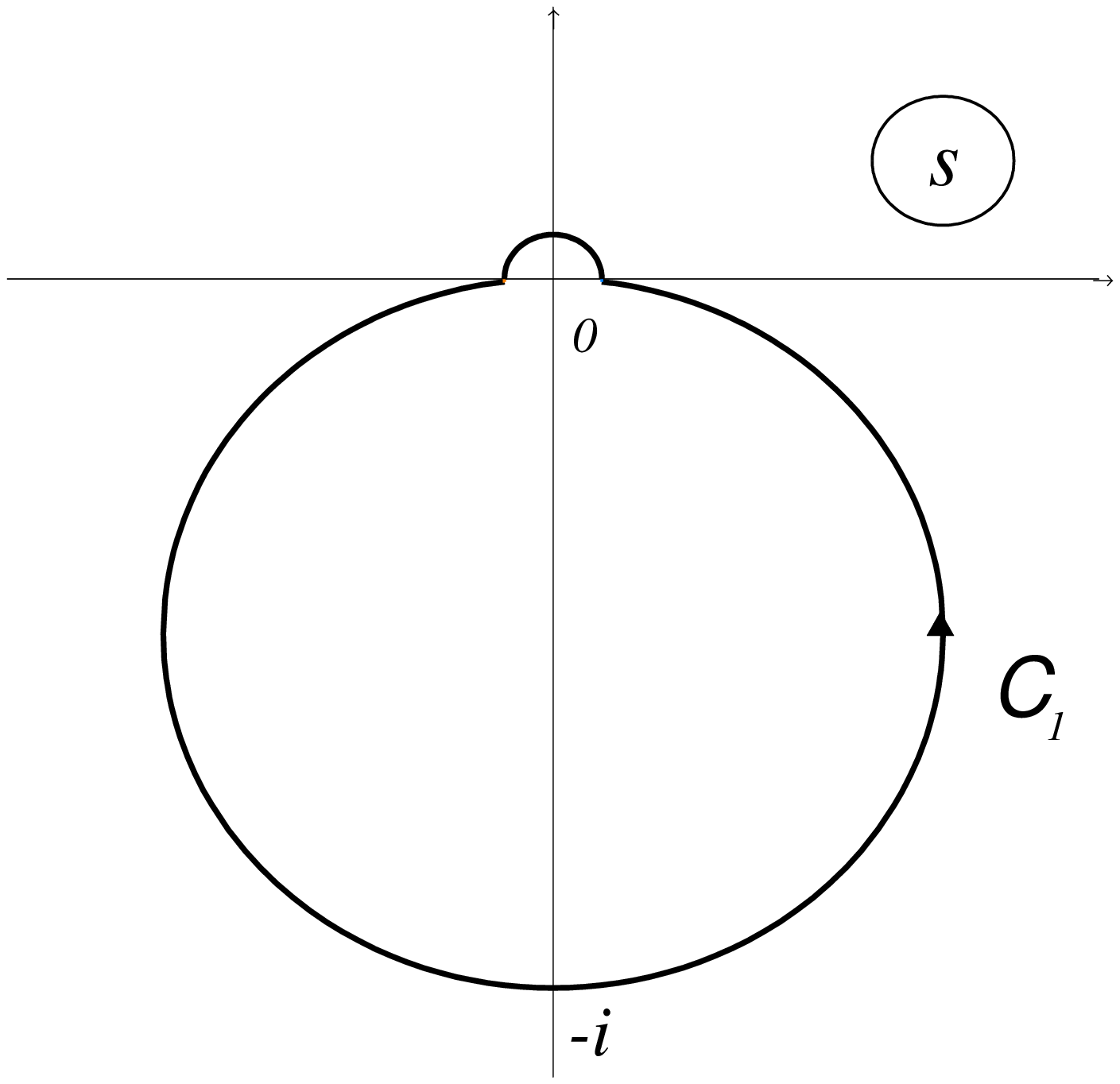,width=1\textwidth}} \caption{The
integration contour $\mathcal{C}_1$ in (\ref{I2}).}
\end{figure*}

\newpage
\begin{figure*}[ht]
\centerline{\psfig{figure=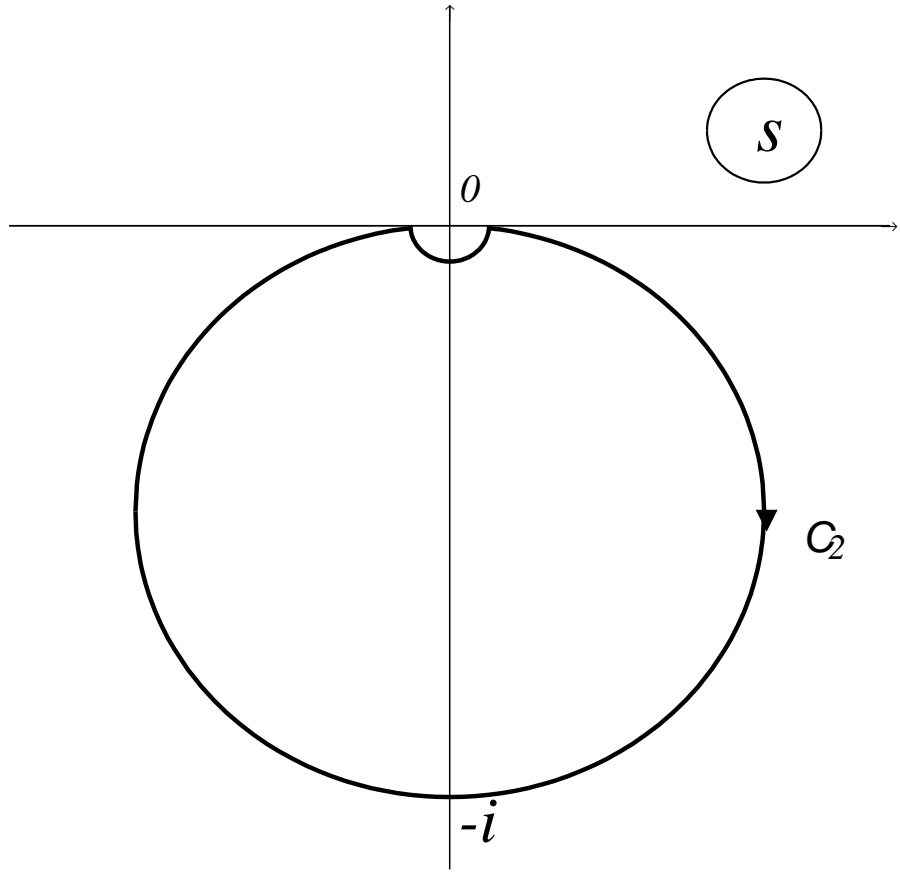,width=1\textwidth}} \caption{The
integration contour $\mathcal{C}_2$ in (\ref{I5}).}
\end{figure*}

\newpage
\begin{figure*}[ht]
\centerline{\psfig{figure=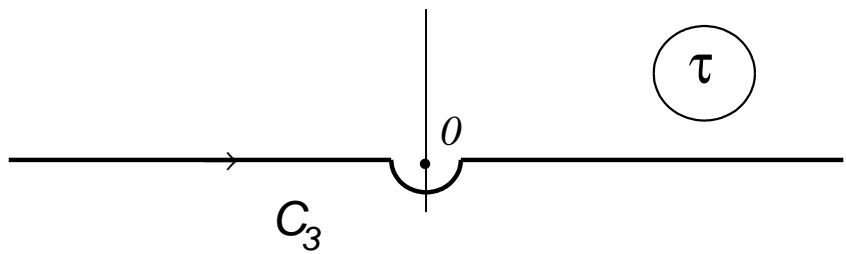,width=1\textwidth}} \caption{The
path of integration $\mathcal{C}_3$ of the convolution integral in
(\ref{Conv3}.)}
\end{figure*}

\end{document}